\newcommand{\fuvcenter}{1530\AA}
\newcommand{\nuvcenter}{2310\AA}

\newcommand{\fuvmag}{\ifmmode{FUV}\else{\it FUV~}}
\newcommand{\nuvmag}{\ifmmode{NUV}\else{\it NUV~}}

\documentclass[12pt]{aastex}
\shorttitle{UV Galaxy Luminosity Function}
\shortauthors{Wyder et al.}

\received{2004 May 5}
\begin{document}
\title{The UV Galaxy Luminosity Function in the Local Universe from GALEX Data}

\author{
Ted K. Wyder\altaffilmark{1},
Marie A. Treyer\altaffilmark{1,2},
Bruno Milliard\altaffilmark{2},
David Schiminovich\altaffilmark{1,3},
St\'ephane Arnouts\altaffilmark{2},
Tam\'as Budav\'ari\altaffilmark{4},
Tom A. Barlow\altaffilmark{1},
Luciana Bianchi\altaffilmark{5},
Yong-Ik Byun\altaffilmark{6},
Jos\'e Donas\altaffilmark{2},
Karl Forster\altaffilmark{1},
Peter G. Friedman\altaffilmark{1},
Timothy M. Heckman\altaffilmark{4},
Patrick N. Jelinsky\altaffilmark{7},
Young-Wook Lee\altaffilmark{6},
Barry F. Madore\altaffilmark{8},
Roger F. Malina\altaffilmark{2},
D. Christopher Martin\altaffilmark{1},
Patrick Morrissey\altaffilmark{1},
Susan G. Neff\altaffilmark{9},
R. Michael Rich\altaffilmark{10},
Oswald H. W. Siegmund\altaffilmark{7},
Todd Small\altaffilmark{1},
Alex S. Szalay\altaffilmark{4}, and
Barry Y. Welsh\altaffilmark{7}}

\altaffiltext{1}{California Institute of Technology, MC 405-47, 1200 East
California Boulevard, Pasadena, CA 91125; wyder@srl.caltech.edu}
\altaffiltext{2}{Laboratoire d'Astrophysique de Marseille, BP 8, Traverse
du Siphon, 13376 Marseille Cedex 12, France}
\altaffiltext{3}{Department of Astronomy, Columbia University, MC2457, 550 W. 120 St., New York, NY, 10027}
\altaffiltext{4}{Department of Physics and Astronomy, The Johns Hopkins
University, Homewood Campus, Baltimore, MD 21218}
\altaffiltext{5}{Center for Astrophysical Sciences, The Johns Hopkins
University, 3400 N. Charles St., Baltimore, MD 21218}
\altaffiltext{6}{Center for Space Astrophysics, Yonsei University, Seoul
120-749, Korea}
\altaffiltext{7}{Space Sciences Laboratory, University of California at
Berkeley, 601 Campbell Hall, Berkeley, CA 94720}
\altaffiltext{8}{Observatories of the Carnegie Institution of Washington,
813 Santa Barbara St., Pasadena, CA 91101}
\altaffiltext{9}{Laboratory for Astronomy and Solar Physics, NASA Goddard
Space Flight Center, Greenbelt, MD 20771}
\altaffiltext{10}{Department of Physics and Astronomy, University of
California, Los Angeles, CA 90095}

\begin{abstract}

We present the results of a determination of the galaxy luminosity
function at ultraviolet wavelengths at redshifts of $z=0.0-0.1$ from
GALEX data.  We determined the luminosity function in the GALEX FUV
and NUV bands from a sample of galaxies with  UV magnitudes between 17
and 20 that are drawn from a total of 56.73
deg$^2$ of GALEX fields overlapping the $b_j$-selected 2dF Galaxy
Redshift Survey. The resulting luminosity functions are fainter than
previous UV estimates and result in total UV luminosity densities of
$10^{25.55\pm0.12}~{\rm ergs~s^{-1}~Hz^{-1}~Mpc^{-3}}$ and
$10^{25.72\pm0.12}~{\rm ergs~s^{-1}~Hz^{-1}~Mpc^{-3}}$ at
\fuvcenter~and \nuvcenter, respectively.
This corresponds to a local star formation rate density in agreement
with previous estimates made with ${\rm H\alpha}$-selected data for
reasonable assumptions about the  UV extinction.

\end{abstract}


\section{Introduction}

In the past few years determinations of the star formation history of
the universe have allowed us to begin to understand quantitatively
when and how the stars in the universe were formed.  Measurements  of
the rest-frame ultraviolet luminosities of galaxies have been
particularly useful in this endeavor.  In the very local universe,
there is a relative lack of systematic surveys of galaxies in the UV.
Before the launch of GALEX, the most comprehensive survey of galaxies
in the local universe was from the FOCA experiment \citep{milliard92},
a balloon-borne telescope that made measurements in a single band
centered at 2000\AA.  Based upon FOCA observations of a total of
$\sim2.2$ deg$^2$, \citet{treyer98} and
\citet{sullivan00} measured the first UV
luminosity function (LF) for a sample of 273 galaxies with
spectroscopic redshifts at $\bar{z}=0.15$.  Their LF has a steep faint
end slope and  a total UV luminosity density, and corresponding star
formation rate density, larger than most previous estimates.  This
higher local UV luminosity density in conjunction with  measurements
at larger distances lead \citet{wilson02} to infer a luminosity
density evolution proportional to $(1+z)^{1.7\pm1.0}$, a trend
shallower than had been estimated previously from the CFRS sample
\citep{lilly96}.

In this letter we present the first results regarding the UV LF based
upon measurements from the {\it Galaxy Evolution Explorer} (GALEX) in
conjunction with redshifts from the 2dF Galaxy Redshift Survey (2dFGRS)
\citep{colless01}.  The new GALEX data allow us to expand upon the
previous FOCA results using a much larger sample drawn from an area of
56.73 deg$^2$ although  to a shallower limiting magnitude of
$m_{UV}=20$.  Throughout this paper, we assume $H_0=70~{\rm
km~s^{-1}~Mpc^{-1}}$, $\Omega_{M}=0.3$ and $\Omega_{\Lambda}=0.7$.

\section{Data}

The data analyzed in this paper consist of 133 GALEX All-Sky Survey
(AIS) pointings that overlap the 2dF Galaxy Redshift Survey in the
South Galactic Pole region. 
The GALEX field-of-view is circular with diameter of $1.2\arcdeg$ and each
pointing is imaged simultaneously in both the FUV and NUV bands
with  effective wavelengths of \fuvcenter~and \nuvcenter,
respectively.  The median exposure time for the fields is 105 seconds,
allowing us to reach a S/N ratio of $\sim5$ for $FUV\approx20.0$ and
$NUV\approx20.5$.  See \citet{martin04} and \citet{morissey04}
for details regarding the GALEX instruments and mission.

Sources were detected and measured from the GALEX images using the
program SExtractor \citep{bertin96}.  As the NUV images are
substantially deeper than the FUV, we used the NUV images for
detection and measured the FUV flux in the same aperture as for the
NUV. The fields analyzed here were processed using a larger SExtractor
deblending parameter DEBLEND\_MINCONT as the standard GALEX pipeline
processing tends to break apart well-resolved galaxies into more than
one source.  We elected to use the MAG\_AUTO magnitudes measured by
SExtractor through an elliptical aperture whose semi-major axis is
scaled to 2.5 times the first moment of the object's radial profile,
as first suggested by \citet{kron80}. All of the apparent magnitudes
were corrected for  foreground extinction using the \citet{schlegel98}
reddening maps and assuming the extinction law of
\citet{cardelli89}. The ratio of the extinction in the GALEX bands to
the reddening $E(B-V)$ was calculated by averaging the extinction law
over each GALEX bandpass, resulting in $A_{FUV}/E(B-V) = 8.376$ and
$A_{NUV}/E(B-V)=8.741$. The median extinction correction for the
galaxies in our South Galactic Pole 
sample is $0.15$ mag in both bands,  with the
corrections ranging from 0.1 to 0.3 mag.

The GALEX catalogs were matched with the 2dFGRS input catalog using a
search radius of $6\arcsec$.  To remove any overlap between adjacent
pointings, we only included sources detected within the inner
$0.45\arcdeg$ of each field.  In addition, sources likely contaminated
by artifacts from bright stars, with 2dF redshift quality flag less
than three or with effective exposure times less than 60 sec were
removed.  Finally, we excluded GALEX sources in regions where the 2dF
redshift completeness was less than 80\%.  After applying all of these
cuts to each band, the total area on the sky of GALEX-2dF overlap is
56.73 deg$^{2}$.

\begin{figure}
\plotone{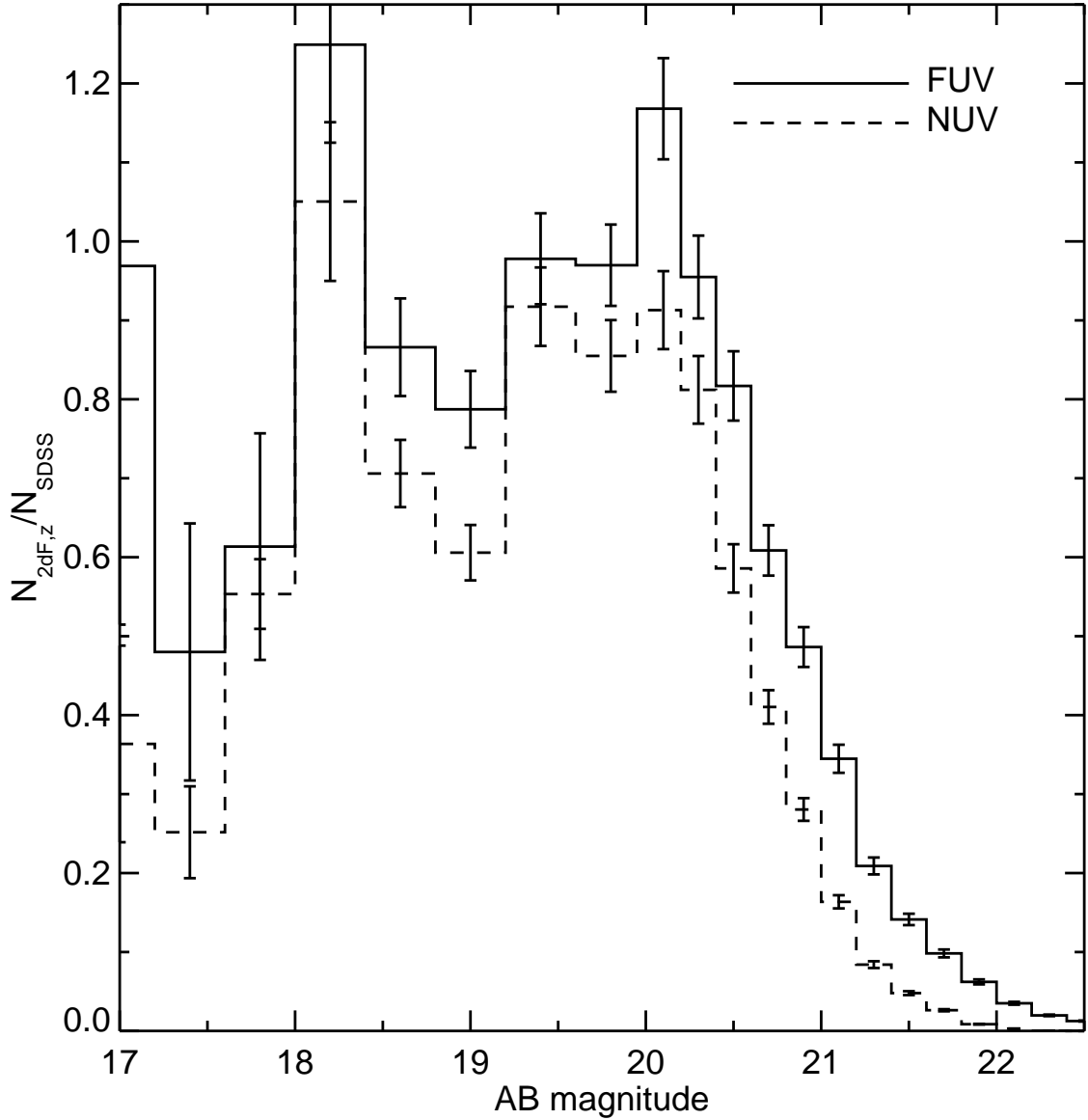}
\caption{Completeness of the GALEX-2dF redshift sample defined as the
ratio of the number counts of galaxies with 2dF redshifts to
the number counts of galaxies as derived from GALEX observations that overlap
the SDSS survey \citep{xu04}.
The solid and dashed lines indicate the FUV and NUV redshift completeness,
respectively.\label{completeness}}
\end{figure}

The GALEX resolution of $6-7\arcsec$ (FWHM) \citep{morissey04} is not
sufficient to accurately separate stars and galaxies. Furthermore, the
2dFGRS input catalog available from the 2dFGRS web
page\footnote{http://www.mso.anu.edu/2dFGRS/}
only includes galaxies brighter than $b_j=19.45$ and
does not include stars.  In order to asses the total completeness of
our 2dF-GALEX matched sample, we normalized our results to the total galaxy
number counts determined by \citet{xu04} based primarily upon 22.64
deg$^{2}$ of GALEX Medium Imaging Survey data overlapping the Sloan
Digital Sky Survey (SDSS) Data Release 1 \citep{abazajian03}. As the
SDSS data include stars and galaxies and reach fainter magnitudes,
they result in a more accurate determination of the galaxy number counts
in the UV.
If we assume that the average galaxy number counts in the SDSS North
Galactic Pole fields are the same as in the GALEX-2dF overlap, then
the redshift completeness of the 2dF matched catalog is given by the
number counts of galaxies with redshifts from 2dF divided by the total
galaxy number counts from the SDSS overlap.  This ratio is shown in Figure
\ref{completeness}.  

\begin{figure}
\plotone{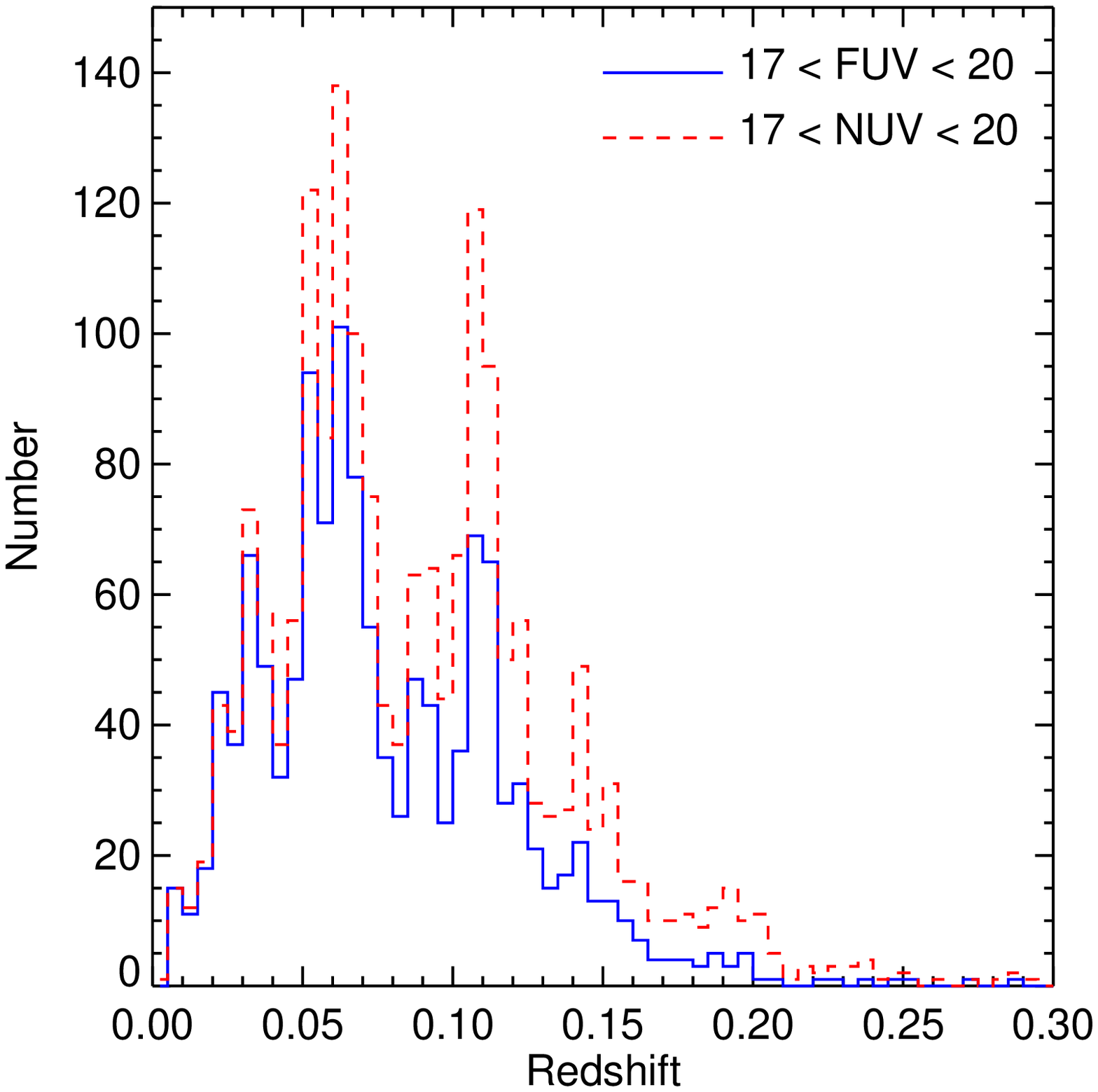}
\caption{The redshift distributions 
of the FUV and NUV selected samples (blue solid and red dashed lines,
respectively) in the range $17\le m_{uv} \le 20$. 
\label{distributions}} 
\end{figure}

The completeness turns over at 20th mag because the redshift sample
becomes incomplete for galaxies with blue $(FUV - b_j)$ or  $(NUV -
b_j)$ colors.  We have limited our LF determination in each band to
galaxies brighter than this limit. To avoid problems with photometry
of large bright galaxies, we also imposed a bright magnitude limit of
17.  The average completeness weighted by the number counts in the
range $17-20$ mag is 92\% in the FUV and 79\% in the NUV.  For objects
with magnitudes brighter than 20.0 in either band, we visually
inspected all of the 2dF spectra and removed a total of 27 objects
with very broad emission lines indicating that the objects are most
likely some sort of AGN.  The redshift distributions for the FUV and
NUV samples are shown in Figure \ref{distributions}.

We further restricted our sample to those galaxies with redshifts $z<0.1$
to insure that our sample is not sensitive to evolution.
The average redshifts are 0.055 and
0.058 in the FUV and NUV, respectively.
After applying all of the cuts mentioned in this section,
a total of 896 galaxies in the FUV and 1124 galaxies in the
NUV remained. The luminosity functions for
the objects with $z>0.1$ are presented in \citet{treyer04}.

\section{Luminosity Functions}

Using the FUV,
NUV and $b_j$ magnitudes, we assigned a best-fit spectral type
to each galaxy using a representative subset of the SEDs from
\citet{bruzual03} and determined
the K-correction needed to transform the observed UV magnitudes to
rest-frame measurements at $z=0$. The K-corrections are in general
quite small ($\lesssim 0.2$).

\begin{figure}
\plotone{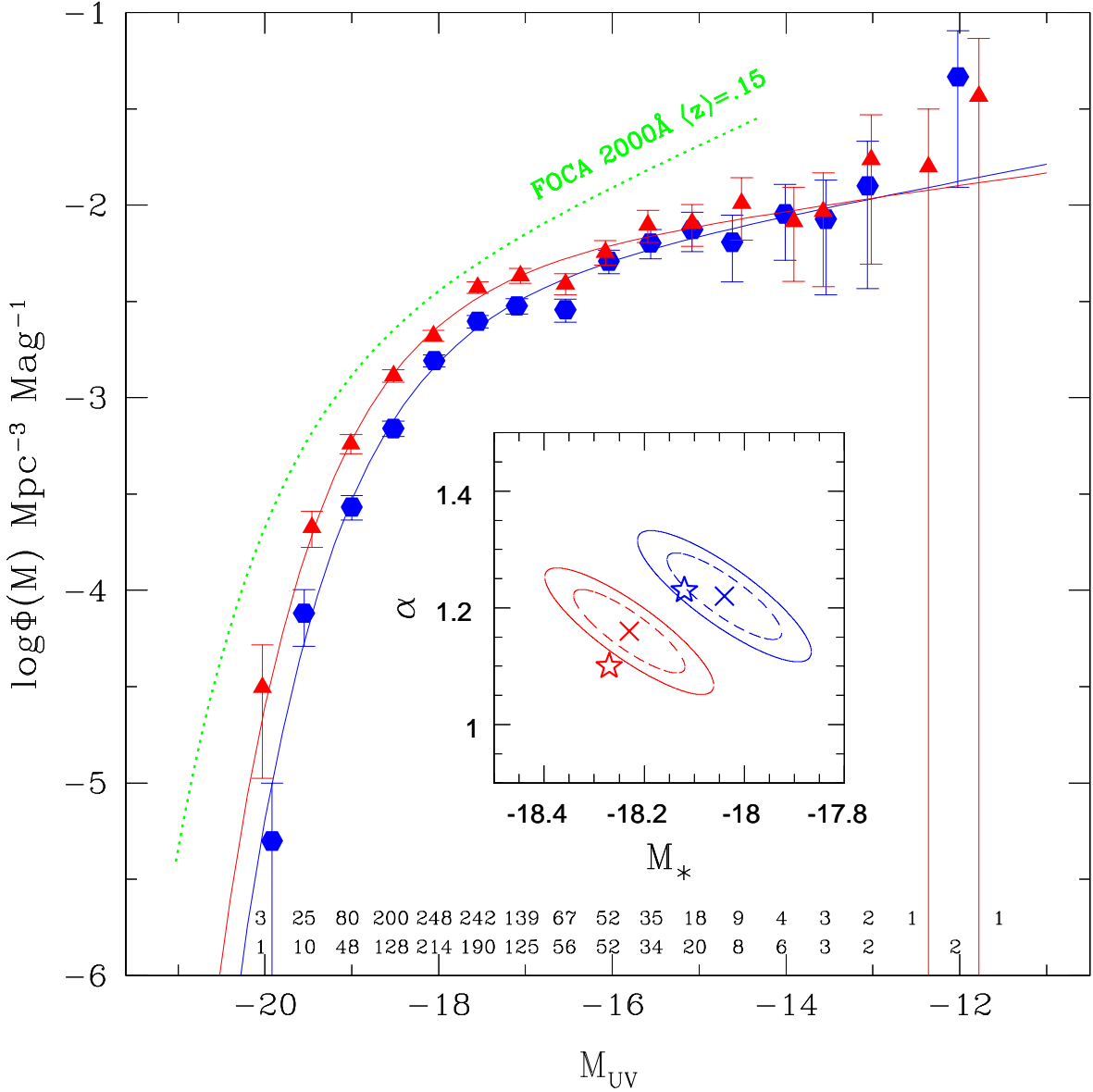}
\caption{The FUV (blue circles) and NUV (red triangles) LFs
for $z<0.1$. The solid lines are the Schechter function fits with best-fit
parameters from Table \ref{params}. The dotted green line
shows the LF measured at ${\rm 2000~\AA}$ from FOCA
data by \citet{sullivan00} over the range of absolute magnitudes
explored in that study. The inset plots the $1\sigma$ error contours
of the Schechter function fits projected into the $M^{\ast} - \alpha$ plane
for the FUV (blue) and NUV (red). The dashed contour shows values with
$\chi^2 - \chi^2_{min} = 1.0$ while the solid contour delineates 
$\chi^2 - \chi^2_{min} = 2.3$ which corresponds to the joint $1\sigma$
uncertainty on $M^{\ast}$ and $\alpha$. The red and blue stars indicate
the best-fitting values of $M^{\ast}$ and $\alpha$ obtained from the
STY method. \label{lf}}
\end{figure}

We determined the LF $\Phi(M)$ and its error
$\sigma(\Phi(M))$ in each band using the $V_{max}$
method \citep{felten76}:
\begin{equation}
\Phi(M) = \sum{f(m)/V_{max}}
\end{equation}
\begin{equation}
\sigma(\Phi(M)) = \left(\sum{f^2(m)/V_{max}^2}\right)^{1/2}
\end{equation}
where $f(m)$ is the inverse of the redshift completeness as estimated in 
\S2 above and $V_{max}$ is the maximum co-moving volume within which
each galaxy
could have been observed given the bright and faint limiting magnitudes
of our sample and its best-fit SED. The resulting LFs
are shown in Figure \ref{lf}.

By minimizing $\chi^2$, we fit the $V_{max}$ LF points
in each band with a Schechter function \citep{schechter76}: 
$\Phi(L)dL=\phi^{\ast} (L/L^{\ast})^{\alpha} e^{-L/L^{\ast}} dL/L^{\ast}$
where $\phi^{\ast}$, $M^{\ast}$ and $\alpha$ were free parameters. The best fit
parameters and their errors, calculated using the range of solutions
within 1.0 of the minimum $\chi^2$, are listed in Table \ref{params} along
with the best-fit LF from \citet{sullivan00} converted
to the AB magnitude system and to $H_0=70$. The errors in $\alpha$
and $M^{\ast}$ are highly correlated and the inset of Figure \ref{lf}
shows the $1\sigma$ error contours projected into the $M^{\ast}-\alpha$
plane. Since the $V_{max}$ method
can be biased in the presence of clustering, we also computed the best-fit
Schechter parameters using the maximum likelihood STY method
\citep{sandage79}. The resulting STY values are listed in Table \ref{params}
and are also plotted in the inset of Figure \ref{lf}. The STY values
lie just inside and outside of the $1\sigma$ $V_{max}$ error ellipses 
in the FUV and NUV, respectively. We adopt the $V_{max}$ results
in the discussion below.

\begin{deluxetable}{lcccccccc}
\rotate
\tablecolumns{8}
\tablewidth{0pc}
\tablecaption{Schechter Function Parameters\label{params}}
\tablehead{
\colhead{} & \colhead{} & \multicolumn{4}{c}{$V_{max}$ method} & \colhead{} &
\multicolumn{2}{c}{STY method} \\
\cline{3-6} \cline{8-9} \\
\colhead{band} & \colhead{$z$}  &  \colhead{$M^{\ast}$} & \colhead{$\alpha$} & 
\colhead{$\log{\phi^{\ast}}$ (Mpc$^{-3}$)} & 
\colhead{$\log{\rho_L}$ (ergs s$^{-1}$ Hz$^{-1}$ Mpc$^{-3}$)} & \colhead{} &
\colhead{$M^{\ast}$} & \colhead{$\alpha$}
}
\startdata
FUV & $0-0.1$ & $-18.04 \pm 0.11$ & $-1.22 \pm 0.07$ & $-2.37 \pm 0.06$ & $25.55 \pm 0.12$ & & $-18.12$ & $-1.23$ \\
NUV & $0-0.1$ & $-18.23 \pm 0.11$ & $-1.16 \pm 0.07$ & $-2.26 \pm 0.06$ & $25.72 \pm 0.12$ & & $-18.27$ & $-1.10$ \\
FOCA & 0.15 & $-19.10 \pm 0.13$ & $-1.51 \pm 0.10$ & $-2.48 \pm 0.11$ & $26.06 \pm 0.15$ & & \nodata & \nodata \\
\enddata
\end{deluxetable}

\section{Discussion}

As can be seen in Figure \ref{lf}, there are significant differences
between the results presented here and those from
\citet{sullivan00}.  The GALEX results have a fainter $M^{\ast}$ in both bands
and have a shallower faint end slope.  The FOCA passband is centered
at 2015\AA~with FWHM of 188\AA~and thus one would expect the FOCA
results to lie in between the GALEX FUV and NUV data.  However, the
FOCA sample is truly an UV-selected sample while that  presented here
relies upon the $b_j$-selected 2dFGRS. This selection could
introduce a bias in our results if the galaxies for which we do not
have redshifts have a different redshift distribution than the
galaxies which are included in our sample. On the other hand, it is
now well established that the UV luminosity density increases with
redshift \citep[e.g.][]{somerville01} and part of the difference is
likely a real effect \citep{treyer04}. However,  the difference of
$\sim0.9$ mag between the FOCA and NUV values for $M^{\ast}$ would
require evolution much larger than determined from other surveys as
well as GALEX data at higher redshifts
\citep{arnouts04,schiminovich04}. A preliminary comparison of the
GALEX and FOCA photometry in a couple of overlapping fields indicates
that the FOCA magnitudes are on average brighter with the difference
becoming larger for fainter sources.  It appears likely these offsets
and non-linearities in the FOCA photometry account for a major part of
the difference between the FOCA and GALEX LFs with the remainder
likely due to a combination of galaxy evolution and the FOCA sample
selection.

In Table \ref{params} we also list the total luminosity density
calculated from the best-fit Schechter parameters
as $\rho_L = \int_{0}^{\infty}{L \Phi(L) dL} = \phi^{\ast} L^{\ast} \Gamma(\alpha+2)$. 
The statistical errors in $\log{\rho_L}$ that take into account the
covariance between the three Schechter function parameters are
0.02 in each band. In addition to this error, 
the uncertainty in the GALEX photometric zeropoint
is $\approx 10\%$ in both bands, corresponding to an uncertainty in
$\log{\rho_L}$ of 0.04. A potentially larger source of error is that
due to large scale structure. The variation in the number density of
galaxies $\bar{n}$ in a contiguous volume $V$ is given approximately by
$\delta \bar{n}/\bar{n} \approx (J_3/V)^{1/2}$ \citep{davis82}
where $J_3$ is an integral
over the galaxy 2-point correlation function and has a value of
$\sim10^4~{\rm Mpc^{3}}$ for a correlation function of the
form $\xi(r) = (r/r_0)^{-\gamma}$ with $r_0=7.21 {\rm Mpc}$
and $\gamma=1.67$ \citep{hawkins03}.
The galaxy number counts from \citet{xu04}
used to set the normalization of our LFs were derived
from approximately 22.64 deg${^2}$.  For $z<0.1$, the corresponding
rms variation in the number density would be $\delta n/n \approx
0.24$, or an uncertainty in  $\delta \log{\rho_L}\approx 0.11$.
Since UV-selected, star-forming galaxies are likely less clustered than
optically-selected
samples, this value is really  an upper limit. Adding these uncertainties
due to large scale structure and calibration in quadrature to the
statistical errors results in a total
uncertainty of $\delta \log{\rho_L} \approx 0.12$ in both bands. 

The spectral slope $\beta$, defined as $f_{\lambda} \propto \lambda^{\beta}$
with $f_{\lambda}$ in units of 
${\rm ergs~s^{-1}~\AA^{-1}~Mpc^{-3}}$,
corresponding to our two luminosity density measurements is
$\beta \approx -1.1$. This is slightly bluer than the slope
of $\beta = -0.9$
determined by \citet{cowie99} at $z=0.7-1.3$ from measurements
at longer rest frame wavelengths spanning ${\rm 1700\AA}$ to
${\rm 2750\AA}$. 

The FUV luminosity density can be used to estimate the star formation
rate (SFR) density in the local universe. For a constant star
formation history and a Salpeter IMF, the SFR is related to the UV
luminosity $L_{\nu}$ (in the range $1500-2800{\rm \AA}$)
by ${\rm SFR~(M_{\sun}~yr^{-1}}) = 
1.4\times10^{-28} L_{\nu} ({\rm ergs~s^{-1}~Hz^{-1}})$
\citep{kennicutt98}. For the FUV luminosity density in Table \ref{params},
we obtain $\log{{\rm (SFR_{FUV}) (M_{\sun}~yr^{-1}~Mpc^{-3})}}=-2.30\pm0.12$
with no extinction correction. For comparison, the
extinction-corrected ${\rm H\alpha}$ LF at $z \lesssim 0.045$ from
\citet{gallego95} shifted to our assumed Hubble constant corresponds
to $\log{{\rm (SFR_{H\alpha})}}=-1.86\pm0.04$ using the ${\rm H\alpha}$ to SFR
conversion from \citet{kennicutt98}.  Based upon ${\rm H\alpha}$
imaging of a subsample of the galaxies used by
\citet{gallego95}, \citet{perez-gonzalez03} argued that the local
${\rm H\alpha}$ luminosity density is $\sim 60\%$ higher due to
uncertainties in the aperture corrections applied to the spectroscopic
data and corresponds to $\log{{\rm (SFR_{H\alpha})}}=-1.6\pm0.2$. Bringing the
FUV SFR into agreement with this result would require an extinction of
$A_{FUV} \simeq 1.8$.

An average extinction of $A_{FUV} \simeq 1.8$ is consistent with a
simple estimate made using the observed $(FUV-NUV)$ colors. While
there is a well-defined relationship between the UV extinction and the
spectral slope for starburst galaxies, more quiescent galaxies tend to
have less extinction for a given UV slope than would be inferred from
nearby starbursts \citep{bell02}.  In particular \citet{kong04} used
the population synthesis models of \citet{bruzual03} along with the
prescription described in \citet{charlot00} for determining how
starlight is absorbed by dust in galaxies to show that the smaller
extinction in non-starburst galaxies can be explained by variations in
the galaxies' star formation histories.  Based upon a set of Monte Carlo
realizations of these models spanning a range of extinctions, ages and
star formation histories, \citet{kong04} were able to approximate the
dependence of the FUV extinction on the UV spectral slope $\beta$ with
the following formula: $A_{FUV} = 3.87 + 1.87(\beta+0.40\log{b})$
where the variable $b$  parametrizes the star formation history and is
defined as the ratio of current to the past average star formation
rate.  Assuming a constant star formation history ($b=1$), an
extinction of $A_{FUV}=1.8$ is obtained for a spectral slope $\beta =
-1.1$, a value consistent with that measured from the FUV and NUV
luminosity densities. On the other hand, the average $(FUV-NUV)$
color of our FUV-selected sample is 0.14, corresponding to a spectral
slope of $\beta=-1.67$. For this $\beta$ and $b=1$, the \citet{kong04}
formula results in $A_{FUV}=0.7$ mag. This extinction is similar to
the results of \citet{buat04} who found that the average extinction
for a local NUV-selected sample is $A_{FUV} \simeq 1$ mag based
upon the FIR to UV
flux ratio. If an extinction of $A_{FUV} \simeq 1$ is more appropriate
for the UV-selected sample presented
here, then the UV-based star formation rate density would be
$\log{{\rm (SFR_{FUV})}}=-1.9\pm0.1$, a value lower than that from
${\rm H\alpha}$ although still consistent to within the errors.  In reality
the extinction is likely a function of absolute magnitude and future
GALEX papers will address in more detail correcting UV fluxes for
extinction in a more rigorous way.

In the near future we will continue our investigation of the UV luminosity
function in the local universe using GALEX AIS data covering 
${\rm \sim 1000~deg^2}$ of the SDSS. In addition to expanding our sample
to include more galaxies, we will use the SDSS
photometry and spectroscopy to explore the dependence of
UV luminosity on other galaxy characteristics, such as color, surface
brightness, environment, metallicity and stellar mass.

\acknowledgements
GALEX (Galaxy Evolution Explorer) is a NASA Small Explorer, launched
in April 2003.  We gratefully acknowledge NASA's support for
construction, operation, and science analysis for the GALEX mission,
developed in cooperation with the Centre National d'Etudes Spatiales
of France and the Korean Ministry of Science and Technology.





\begin{thebibliography}

\bibitem[Abazajian et al.(2003)]{abazajian03} Abazajian, K. et al. 2003,
\aj, 126, 2081
\bibitem[Arnouts et al.(2004)]{arnouts04} Arnouts, S., et al. 2004, this volume
\bibitem[Bell(2002)]{bell02} Bell, E. F. 2002, \apj, 577, 150
\bibitem[Bertin \& Arnouts(1996)]{bertin96} Bertin, E., \& Arnouts, S.
1996, \aaps, 117, 393
\bibitem[Bruzual \& Charlot(2003)]{bruzual03} Bruzual, G., \& Charlot, S.
2003, \mnras, 344, 1000
\bibitem[Buat et al.(2004)]{buat04} Buat, V., et al. 2004, \apjl, this volume
\bibitem[Cardelli et al.(1989)]{cardelli89} Cardelli, J. A., Clayton, G. C.,
\& Mathis, J. S. 1989, \apj, 345, 245
\bibitem[Charlot \& Fall(2000)]{charlot00} Charlot, S., \& Fall, S. M. 2000,
\apj, 539, 718
\bibitem[Colless et al.(2001)]{colless01} Colless, M., et al. 2001, \mnras,
328, 1039
\bibitem[Cowie et al.(1999)]{cowie99} Cowie, L. L., Songaila, A., \&
Barger, A. J. 1999, \aj, 118, 603
\bibitem[Davis \& Huchra(1982)]{davis82} Davis, M., \& Huchra, J. 1982,
\apj, 254, 437
\bibitem[Felten(1976)]{felten76} Felten, J. E. 1976, \apj, 207, 700
\bibitem[Gallego et al.(1995)]{gallego95} Gallego, J., Zamorano, J.,
Arag\'on-Salamanca, \& Rego, M. 1995, \apj, 455, L1
\bibitem[Hawkins et al.(2003)]{hawkins03} Hawkins, E., et al. 2003, \mnras,
346, 78
\bibitem[Kennicutt(1998)]{kennicutt98} Kennicutt, R. C. 1998, \araa, 36, 189
\bibitem[Kong et al.(2004)]{kong04} Kong, X., Charlot, S., Brinchman, J., \&
Fall, S. M. 2004, \mnras, 349, 769
\bibitem[Kron(1980)]{kron80} Kron, R. G. 1980, \apjs, 43, 305
\bibitem[Lilly et al.(1996)]{lilly96} Lilly, S. J., Le F\`evre, O., 
Hammer, F., \& Crampton, D. 1996, \apj, 460, L1
\bibitem[Martin et al.(2004)]{martin04} Martin, C. et al. 2004, \apjl,
this volume
\bibitem[Milliard et al.(1992)]{milliard92} Milliard, B., Donas, J., 
Laget, M., Armand, C. \& Vuillemin, A., 1992, \aap, 257, 24
\bibitem[Morissey et al.(2004)]{morissey04} Morissey, P. et al. 2004, \apjl,
this volume
\bibitem[P\'erez-Gonz\'alez et al.(2003)]{perez-gonzalez03}
P\'erez-Gonz\'alez, P. G., Zamorano, J., Gallego, J., Arag\'on-Salamanca,
A., \& Gil de Paz, A. 2003, \apj, 591, 827
\bibitem[Sandage et al.(1979)]{sandage79} Sandage, A., Tammann, G. A., \&
Yahil, A. 1979, \apj, 232, 352
\bibitem[Schechter(1976)]{schechter76} Schechter, P. 1976, \apj, 203, 297
\bibitem[Schiminovich et al.(2004)]{schiminovich04} Schiminovich, D., et al. 2004, this volume
\bibitem[Schlegel et al.(1998)]{schlegel98} Schlegel, D. J., Finkbeiner,
D. P., \& David, M. 1998, \apj, 500, 525
\bibitem[Somerville et al.(2001)]{somerville01} Somerville, R. S., 
Primack, J. R., \& Faber, S. M. 2001, \mnras, 320, 504
\bibitem[Sullivan et al.(2000)]{sullivan00} Sullivan, M., Treyer, M. A., 
Ellis, R. S., Bridges, T. J. Milliard, B., \& Donas, Jos\'e 2000, \mnras, 
312, 442
\bibitem[Treyer et al.(1998)]{treyer98} Treyer, M. A., Ellis, R. S., 
Milliard, B., Donas, J., Bridges, T. J. 1998, \mnras, 300, 303
\bibitem[Treyer et al.(2004)]{treyer04} Treyer, M. A., et al. 2004, this volume
\bibitem[Wilson et al.(2002)]{wilson02} Wilson, G., Cowie, L. L., Barger, 
A. J., \& Burke, D. J. 2002, \aj, 124, 1258
\bibitem[Xu et al.(2004)]{xu04} Xu, C. K. et al. 2004, \apjl, this volume

\end{thebibliography}
\end{document}